\documentclass[aps,preprint,a4paper,showpacs,amsmath,amssymb,floatfix]{revtex4}
\usepackage{graphicx}
\usepackage{dcolumn}
\usepackage{bm}
\newcommand{\Op}[1]{{\boldsymbol{\mathrm{\hat{#1}}}}}

\begin{document}
\draft

\title{ Short Time Cycles of Purely Quantum Refrigerators }

\author{Tova Feldmann and Ronnie Kosloff}

\address{
Institute of Chemistry
the Hebrew University, Jerusalem 91904, Israel\\
}

\begin{abstract}
Four stroke Otto refrigerator cycles with 
no classical analogue are studied. Extremely short 
cycle times with respect to the internal time scale of the working medium characterize these refrigerators. 
Therefore,  these cycles are termed {\em sudden}. 
The sudden cycles are characterized by  the stable limit cycle which is the invariant of the global cycle propagator.
During their operation the states of the working medium possess  significant coherence which is not erased 
in the equilibration segments due to the very short time allocated.
This characteristic is reflected in a  difference between the energy 
entropy and the Von Neumann entropy of the working medium.
A classification scheme for sudden refrigerators is developed allowing simple approximations 
for the cooling power and coefficient of performance.
\end{abstract}
\maketitle

\section{Introduction} 
\label{sec:introduction}

Magnetic cooling was initiated more than a half century
ago \cite{kramers1934}. The cooling agent paramagnetic salts was cooled 
both by adiabatic demagnetization \cite{kramers1934} and by adiabatic magnetization 
\cite{wolf1959} depending on the sign of the isothermal change of
entropy  as a function of the magnetic field. 
Currently, adiabatic demagnetization is an efficient technique for cooling detectors in 
space missions and also for  home refrigerators without freon gas \cite{hakonen91,gschneider05,lchen06,fattori06}. 
These devices are realizations of quantum refrigerators 
a subject  which has been of growing interest in the last decade 
\cite{gordon00,feldmann96,feldmann00,kosloff03,feldmann06,rowea06,hensler05,rezeky06,segal06,geva0,heJ09,quan07,rezek09,alla10,popescu10,feldmann2010,alla11,salamon11}. 
In this study we explore an extreme mode of operation of magnetic refrigerators
where the short time allocation on the cycle segments leads to  quantum characteristics.

A prototype of a magnetic four stroke refrigerator is investigated.
The working medium of the refrigerator is a magnetic material modeled as 
an ensemble of two coupled spin systems \cite{kosloff01,thomas11}.  The working cycle is
composed of four segments: two isomagnets, one magnetization
and one demagnetization segment. 
The dynamics is generated by a completely positive map \cite{kraus71}, which settles
in a self repeating cycle - the limit cycle \cite{feldmann04}.

In the  microscopic model two timescales emerge, the cycle time and the internal
timescale determined by frequencies of the working medium.
When the allocated time on each segment of the cycle  is large compared to the internal timescale, 
the cycle is quasi-adiabatic. Under these circumstances the state of the working 
medium is close to an equilibrium or Gibbs state. 
These states are characterized by the expectation value
of the energy. As a result the state $\Op \rho$ is diagonal in the energy 
representation $[\Op \rho, \Op H]=0$.

When the cycle period becomes comparable to the internal timescale, the state 
of the system lag behind the change of the external parameters. 
The cause of this effect is that the external control Hamiltonian does not commute with the internal
Hamiltonian. In this case, the states of the working fluid can not follow the instantaneous
energy levels, therefore additional energy is stored in the working medium.
This additional energy is accompanied by large off-diagonal elements 
of the state of the working medium $\Op \rho$. 
The dissipation of this additional energy is the quantum analogue of friction \cite{kosloff03}.

A quantitative distinction can be made between two sets of cycles depending 
if the time spent on each segment is shorter or longer than the internal timescale  This distinction is a boundary range between the two sets.
The \em sudden \rm cycles are characterized by segment times
which are much shorter than the internal dynamical timescale. 
The  \em regular \rm  
cycles are those with segment times which are longer than the internal timescale.
Our previous studies focused on \em regular \rm  cycles \cite{kosloff10,feldmann2010}. 
Intermediate cycles exists where some segments have a short time allocation  compared to the internal timescale and other segments have a long time allocation.

In the  \em regular \rm cycles the different segments can be characterized  and studied individually. In \cite{kosloff10} Fig. 4 is a representative example of those cycles where the different segments have many cycles, and can be treated almost independently, as opposed to the \em sudden \rm cycles of the recent study.
Typically, the states of the \em regular \rm cycles are almost diagonal in the energy representation.
This energy dominance is even more pronounced at the contact points between the cycle segments.
The optimal cycles of the \em regular \rm type are strictly diagonal in the energy representation 
on the contact points. These cycles are termed frictionless \cite{rezek09,kosloff10,muga09,salamon11}.

The \em sudden \rm cycles are characterized by large off-diagonal elements in the energy representation
$[\Op \rho, \Op H] \ne 0$. 
As a result, the \em sudden \rm cycles are characterized by the Von Neumann entropy 
${\cal S}_{VN}=- tr \{ \Op \rho \ln \Op \rho \}$  being different from the energy entropy 
${\cal S}_E= - \sum p_n \ln p_n$  where $p_i$ the expectation value of the $n$th energy level.
${\cal S}_E \ge {\cal S}_{VN}$ with equality only when $[\Op \rho, \Op H]= 0$.
These cycles are different from the time optimal bang bang type cycles which have significant off-diagonal  
elements in the energy representation on the {\em adiabatic \rm} segments but are diagonal on the contact points \cite{rezek09,kosloff10,muga09,salamon11}.

The large off diagonal elements of the density operator $\Op \rho$
in the \em sudden \rm cycles accompany their global behavior. 
The elements do not vanish on the connecting points between segments.
As a result additional global links are generated. 
The interrelation of the observable values generate off diagonal 
elements of $\Op \rho$.
Increasing further the cycle time,  the amplitude of the off-diagonal parts is 
suppressed. The coherence, defined in Sec. II characterizes  the 
\em suddenness\rm, showing that it decreases with increasing cycle time.

Another important property of the {\em sudden \rm}  cycles is that their 
performance is not necessarily optimal. Starting from an arbitrary initial state
the relaxation toward a limit cycle of sudden cycle might take
several thousands of iterations, as opposed to the {\em regular \rm} cycles,
where generally several iterations are sufficient.
When the  \em sudden \rm cycle is closed, and leads to refrigeration,
there is a close neighborhood of parameters which lead to similar cycles.
These neighborhoods of analytic behavior are very small.  
As a result there are islands of parameters which lead to {\em sudden \rm} 
refrigeration cycles which are disconnected from other islands.

To gain insight on the {\em sudden \rm} cycles, analytical approximate 
expressions for the cooling power of
the cycle, its coefficient of performance (COP), the coherence as a function
of time, and the entropy generation of a cycle are obtained
It is shown, both numerically and by analytic approximations, that 
the cooling power of the {\em sudden \rm} cycles achieve a 
maximum value as a function of the inverse temperature multiplied by the
coefficient of the inner coupling, as opposed to  
{\em regular \rm} cycles which depend exponentially on the inverse temperature.


\section{The Cycle of Operation, the Quantum Heat Pump}
\label{sec:cycle}

A heat pump extracts heat from a cold
reservoir, and transfers it to the hot reservoir.
The operation of the heat pump is determined by the properties
of the working medium and the coupling to the cold and hot baths.
The cycle of operation is defined by
the external controls which include the variation in time of the 
field with the periodic property $\omega(t)=\omega(t+\tau)$
where $\tau$ is the total cycle time synchronized with  the contact times
on the different segments of the cycle.
The cycle studied is composed of the following four segments, 
see Fig. \ref{fig:typical}:

\begin{enumerate}

\item {Segment  $A \rightarrow B$ (termed \em isomagnetic \rm or 
           \em isochore \rm), the field
          is maintained  constant $\omega=\omega_c$, the working medium
       is in contact with the cold bath of temperature $T_c$ with heat
       conductance $\Gamma_c$, and dephasing parameter $\gamma_c$ for a 
       period of $\tau_c$, and with propagator ${\cal U}_c$.}

\item   {Segment ~$B \rightarrow C$ (termed \em magnetization \rm or 
         \em compression adiabat \rm),
        the field  changes from $\omega_c$ to $\omega_h$ in a time period 
        of $\tau_{ch}$ with propagator ${\cal U}_{ch}$.}

\item   {Segment ~~$C \rightarrow D$ \em isomagnetic \rm, 
          or \em isochore \rm, the field is 
        maintained constant $\omega=\omega_h$, the working medium
        is in contact with the hot bath of temperature $T_h$ with
        heat conductance $\Gamma_h$, and dephasing parameter $\gamma_h$ 
        for a period of $\tau_h$ and with propagator ${\cal U}_h $.}

\item   {Segment ~$D \rightarrow A$ \em demagnetization \rm or 
        \em expansion adiabat \rm, the field changes from $\omega_h$ 
          to $\omega_c$ in a time period  of 
         $\tau_{hc}$, with propagator ${\cal U}_{hc}$.}

\end{enumerate}

In the basic paper, \cite{kosloff03}, section III
describes in detail the definitions of the segments \em isochore \rm  and   \em adiabat \rm.
A somewhat different approach can be found in the paper of  Quan et al,  \cite{quan07}.  

At the limit cycle, all the values of the expectation values exactly repeat 
themselves during each cycle time $\tau$. The propagator of the cycle will 
be termed ${\cal U}_{global}$,  where ${\cal U}_{global}$ is constructed 
by the individual propagators (for example at point $A$ 
in  Fig. \ref{fig:typical}), as:
\begin{equation}
{\cal U}_{global}~~=~~        {\cal U}_{hc} {\cal U}_h {\cal U}_{ch} {\cal U}_c      
\label{eq:globalprop}  
\end{equation} 
 The limit cycle is characterized by an invariant eigenvector of ${\cal U}_{global}$, with eigenvalue 1(one).  

The dynamics of the refrigerator's working medium follow our previous studies
\cite{kosloff03,feldmann04,kosloff10}. We construct the segment propagators ${\cal U}$ by solving the dynamics for quantum thermodynamical observables. This dynamics is generated by 
completely positive maps within the formulation of quantum open systems
\cite{lindblad76,alicki87}. It is generated by the Liouville 
superoperator, $ {\cal L}$ in the Heisenberg picture,
\begin{equation}
\frac {d {\Op A}}{dt}~~=~~ i[{\Op H}, {\Op A}]+ {\cal L}_{D}( {\Op A})
~+~ \frac{\partial {\Op A}}{\partial t}~~~.
\label{eq:heisenberg}
\end{equation}  
where ${\cal L}_{D}$ is a generator of a completely positive dissipative
Liouville superoperator, which includes the temperatures $T_{c/h}$ of the 
reservoirs. The state of the system $\Op \rho$ is then reconstructed from a finite set of observables.

The Hamiltonian of the working fluid has the structure:
\begin{equation}
\Op H = \Op H_{int} + \Op H_{ext}(t)    
\label{eq:hamil}  
\end{equation} 
where $ [\Op H_{int} ,\Op H_{ext}(t)] \ne 0$. We chose a cycle that when 
the working fluid is in thermal contact with the reservoirs the Hamiltonian is stationary.

Motivated by a refrigerator operating with magnetic  salt \cite{hakonen91,gschneider05,lchen06,fattori06},
a simple model was constructed with a working fluid composed of pairs of coupled spins. 
The uncontrolled, internal Hamiltonian becomes($\hbar=1$):
${\Op H}_{int} ~~=~~\frac {1} {2}  J \left({ {\boldsymbol{\mathrm{\hat
{\sigma}}}}_x^1} \otimes { {\boldsymbol{\mathrm{\hat{\sigma}}}}_x^2} -
{{\boldsymbol{\mathrm{\hat{\sigma}}}}_y^1}\otimes {\boldsymbol
{\mathrm{\hat{\sigma}}}}_y^2 ~~~
  \right) ~\equiv~J {\Op B_2}$~~~
where ${{\boldsymbol{\mathrm{\hat{\sigma}}}}}$ represents the spin-Pauli 
operators, and $J$ scales the strength of the inter-particle interaction, 
which is assumed to be constant, for a given pump. When $J = 0$, the model 
represents a working medium with noninteracting atoms \cite{feldmann96}.
The external Hamiltonian is chosen to be:
${\Op H}_{ext} ~~=~~\frac {1}{2} \omega(t)
\left({\boldsymbol{\mathrm{\hat{\sigma}}}}_z^1
\otimes {\bf \hat I^2}
+
{\bf \hat I^1} \otimes {{{{\boldsymbol{ \mathrm {  \sigma}}}}}_z^2}
\right)~\equiv~\omega(t) {\Op B_1}$~~~~
where $\omega(t)$ represents the external control field. 
The total Hamiltonian becomes:
${ \bf \hat H } ~~=~~\omega(t) {\bf \hat B_{1}}+\rm J {\bf \hat B_{2}}$

The eigenvalues of $\Op H$ are $ \epsilon_1= - {\Omega}(t) ,~
\epsilon_{2/3}=0,~ \epsilon_4= {\Omega}(t) $ 
where $\Omega(t)=\sqrt{\omega(t)^2+J^2}$, which is the temporary energy scale,
which is dependent on both the internal and external parts of the Hamiltonian. 
At various  times  $\Op H(t)$ does not commute with itself since: 
~~~$[ {\Op B_1}, {\Op B_2}] \equiv  2 i {\Op B_3} \neq 0$ ,~~~~
${\Op B_3}
~~=~~\frac{1}{2}  \left({ {\boldsymbol{\mathrm{\hat
{\sigma}}}}_y^1} \otimes { {\boldsymbol{\mathrm{\hat{\sigma}}}}_x^2} +
{{\boldsymbol{\mathrm{\hat{\sigma}}}}_x^1}\otimes {\boldsymbol
{\mathrm{\hat{\sigma}}}}_y^2 ~~~  \right)$

The explicit equation for the dissipative part ${\cal L}_{D}(\bf \hat A)$ \cite{lindblad76} is
\begin{eqnarray}
{\cal L}_{D} (\bf \hat A)  ~~=~~
\Large 
\sum_{j}~\left(
{\bf \hat F}_{\rm j} 
{\bf \hat A} {\bf \hat F}_{\rm j}^{\dagger} -  \frac{ 1}{ 2 } ( \bf \hat F_{\rm j}
{\bf \hat F}_{\rm j}^{\dagger} {\bf \hat A}~+~{\bf \hat A} {\bf \hat F}_{\rm j}
{\bf \hat F}_{\rm j}^{\dagger}) \right)~~~,
\label{eq:Lindblad0}
\end{eqnarray}  
where the $ (\bf \hat F_{\rm j})$  were identified with the raising and lowering
operators, from energy level $n$ to $n-1$ and vice versa.  
These operators were constructed by first diagonalizing the Hamiltonian, 
then defining the $ (\bf \hat F_{\rm j})$ operators  in the
energy representation. The lowering transition rates $\kappa_{c/h}^{\downarrow}$
were chosen to be equal for all the four transitions, while
the raising transition rates  $\kappa_{c/h}^{\uparrow}$ were obtained by
 forcing detailed balance relation on the hot and cold isomagnetic segments
 $\kappa_{c/h}^{\uparrow} / \kappa_{c/h}^{\downarrow}   ~=~\exp(-\Omega_{c/h} / T_{c/h})$.
The heat transfer rate to the baths will be 
$\Gamma_{c/h}=\kappa_{c/h}^{\downarrow} ~+~\kappa_{c/h}^{\uparrow}, (k_B=1) $
\cite{kosloff03}, see also the Appendix.

In addition a dissipative generator of elastic encounters is added described as 
\begin{equation}
{\cal L}_{D^e}( {\Op A})~~=~~ - \gamma  [{\Op H},[ {\Op H}, {\Op A}]]
\label{eq:dephasing}
\end{equation} 
where $\gamma$ is the dephasing constant.
  
The state of the system $\Op \rho$ can be expanded by a complete set of orthogonal operators 
on the Hilbert space of the system. The expansion coefficients are proportional to the 
expectation values of these operators. This set  can form a complete vector space to represent 
the propagators  $ {\cal U}_i $ on different segments. 

A  thermodynamically inspired set of observables is a minimum set of operators which completely 
defines the state of the working medium when it reaches the limit cycle. 
For thermal equilibrium the energy and identity operators are sufficient. For the limit cycle
this set has to be expanded.
The set is initiated from the energy $\Op H$ and new operators are added 
which are dynamically coupled to the energy.
This set is formed
from a linear combinations of the stationary closed set ~$ \left\{ \Op B \right\} $ of  operators: 
\begin{equation}
{\Op H} ~ ~=~\omega (t) {\Op B_1 } ~+~J {\Op B_2 }
~~
 {\Op L} ~~=~-J {\Op B_1 } ~+~\omega(t) {\Op B_2 } 
~~
 {\Op C}  ~ ~=  \Omega(t)     {\Op B_3 } 
\label{defnew}   
\end{equation}
The three operators defined in Eq. (\ref{defnew}) form a closed Lie algebra, 
for they are linear combinations of the original operators  
$ {\bf \hat B_i} $, which also form a closed Lie algebra.
 
To uniquely define the diagonal part of the state $\Op \rho$ in the energy representation, 
the original set has to be supplemented with two additional operators:
${\Op V} = \Omega {\Op B_4}= \frac{1}{2}\Omega ({\Op I}^1 \otimes {\Op \sigma}_z^2 -{\Op I}^2 \otimes {\Op \sigma}_z^1)$ and 
${\Op D} = \Omega {\Op B_5}=\Omega {\Op \sigma}_z^1 \otimes {\Op \sigma}_z^2$. With this operator base the state $\Op \rho$ can be expanded as:
\begin{equation}
\Op \rho =\frac{1}{4} {\Op I} + \frac{1}{ \Omega} \left( \langle \Op H \rangle \Op H +\langle \Op L \rangle \Op L 
+\langle \Op C \rangle \Op C  + \langle \Op V \rangle \Op V +\langle \Op D \rangle \Op D  \right) 
\label{eq:rho}
\end{equation}
$\Op V$ and $\Op D$ commute with $\Op H$. The equilibrium value of $\langle \Op V \rangle$ is  zero,
and once it reaches equilibrium it does not change during the cycle dynamics. This means that on the limit cycle
the state $\Op \rho $ can be reconstructed by four expectation values:
$E = \langle \Op H \rangle $,  
$L = \langle \Op L \rangle $, $C = \langle \Op C \rangle $ and $D = \langle \Op D \rangle $.
In the energy representation the state $\Op \rho$ becomes:
\begin{eqnarray}
\Op \rho_e = \frac{1}{4} \left(
\begin{array}{cccc}
1+ \frac{1}{\Omega}(D -2E)&0&0&\frac{2}{\Omega}(L+iC)\\
0&1- \frac{1}{\Omega}D &0&0\\
0&0&1- \frac{1}{\Omega}D &0\\
\frac{2}{\Omega}(L-iC)&0&0&1+ \frac{1}{\Omega}(D+2E)
\end{array}
\right)
\label{eq:rhoe} 
\end{eqnarray}

A measure of the off diagonal elements in the energy frame is the  coherence \cite{k108}:
\begin{equation}
\tilde {\cal C} = tr \{ (\Op \rho_e - \Op \rho_{ed})^2\}~,
\label{eq:coherence}  
\end{equation}
where $\Op \rho_{ed}$ is the diagonal stationary part of the density operator in the energy frame. 
>From Eq. (\ref{eq:rhoe}),~ the coherence becomes $\tilde {\cal C}= \frac {L^2+C^2} {2 \Omega^2}$.
Fig. \ref{fig:fourcoherence} shows the transition from a sudden cycle to a regular  one on the 
\em adiabatic \rm segment. 
An increase in the time allocation on this segment decrease the coherence $\tilde {\cal C}$.

The vector space defining the propagators is constructed from the four operators
$\Op H,\Op L,\Op C , \Op D$ and the identity $\Op I$ $\vec X =(\Op H,\Op L,\Op C , \Op D,\Op I)$.
Using this set the propagator $ {\cal U}_{i}(\tau)$  on the \em isochores \rm 
(or equivalently on the \em isomagnets) \rm becomes \cite{kosloff03};
\begin{eqnarray}
{\cal U}_{i}=
\left(
\begin{array}{ccccc}
e^{(-\Gamma_{c/h} \tau)}  & 0 & 0& 0  & {E}_{eq}
(1-e^{(-\Gamma_{c/h} \tau)}) \\
0  &  Kcos(\Omega \tau)      & -Ksin(\Omega \tau) & 0 & 0  \\
0  &  Ksin(\Omega \tau) & Kcos(\Omega \tau) & 0 & 0 \\ 
\frac {1}{\Omega}(E_{eq}(e^{-\Gamma_{c/h}\tau}-e^{-2\Gamma_{c/h}\tau}))   &  0 & 0 & e^{-2\Gamma_{c/h} \tau}  &- \frac{E_{eq}^{2}}{\Omega}(e^{-\Gamma_{c/h} \tau}-1) \\
0  &  0   & 0  & 0 & 1  \\
\end{array}
\right)
\label{eq:propiso}  
\end{eqnarray}   
where $ K= e^{(-[\Gamma_{c/h}+\gamma_{c/h} \Omega^2] \tau )},
~~ \Gamma_{c/h},~ \gamma_{c/h},~\Omega, ~\tau$ are defined above. 

The periodic functions in Eq. (\ref{eq:propiso})  mean that the isomagnetic segments
are quantized.  Whenever $\Omega \tau~=~2 \pi$, the two coupled equations
of $ L,C$  complete a period. The other two
expectation values $E$ and $D$ are decoupled from $L,C$ the expectation 
values of the operators $\Op L$ and $\Op C$. 
Quantization exists also on the \em adiabats \rm.  
Closed form solutions for the  \em adiabats \rm leading to quantized motion are obtaind for a constant adiabatic parameter $\mu =  \frac {J \dot \omega}{\Omega^3} $ \cite{kosloff10}.
\begin{figure}[tb]
\vspace{1.2cm} 
\center{\includegraphics[height=5cm]{cyc430atA.eps}}
\center{\includegraphics[height=5cm]{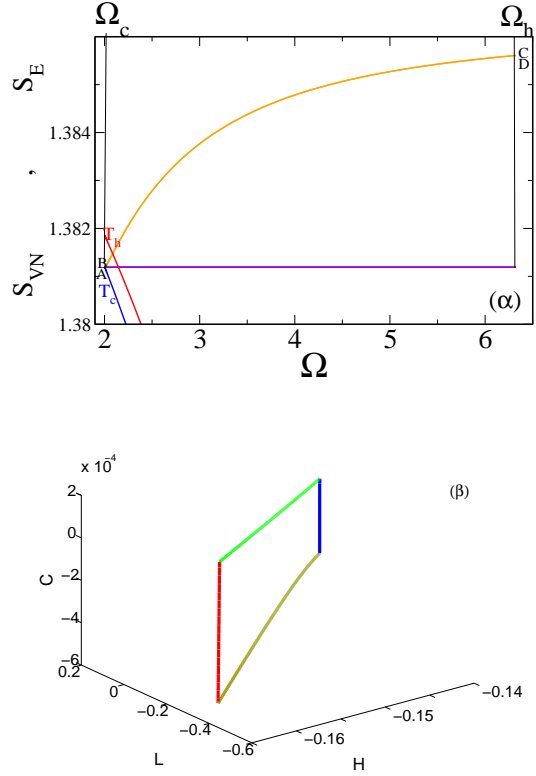}}
\vspace{0.1cm} 
\caption{(Color online)Top($\alpha$): An example of an extreme sudden cycle 
with analytical scheduling, in the ($\Omega, S_E$) plane (curved shape ) and ($\Omega, S_{VN}$) plane (shrinked rectangle shape), together with the isotherms corresponding to the cold/hot baths temperatures, 
$T_c/T_h$. The cooling power $Q_c/\tau=1.2 \cdot 10^{-6}$.
The cyle parameters are: $T_c=14, T_h=15, J=2., \omega_c=0.1$, $\omega_h=6.$.
The time allocations: $\tau_{c},\tau_{ch}, \tau_h,\tau_{hc}=0.9,0.00035,0.00025,0.00035, \kappa_h^{\downarrow}=0.36,~\kappa_c^{\downarrow}=0.328 $,
Bottom($\beta$): the corresponding cycle trajectory in the $H,L,C$ space.}
\label{fig:typical2}     
\end{figure}

\begin{figure}[tb]
\vspace{1.2cm} 
\center{\includegraphics[height=5cm]{cyctypical.eps}} 
\center{\includegraphics[height=5cm]{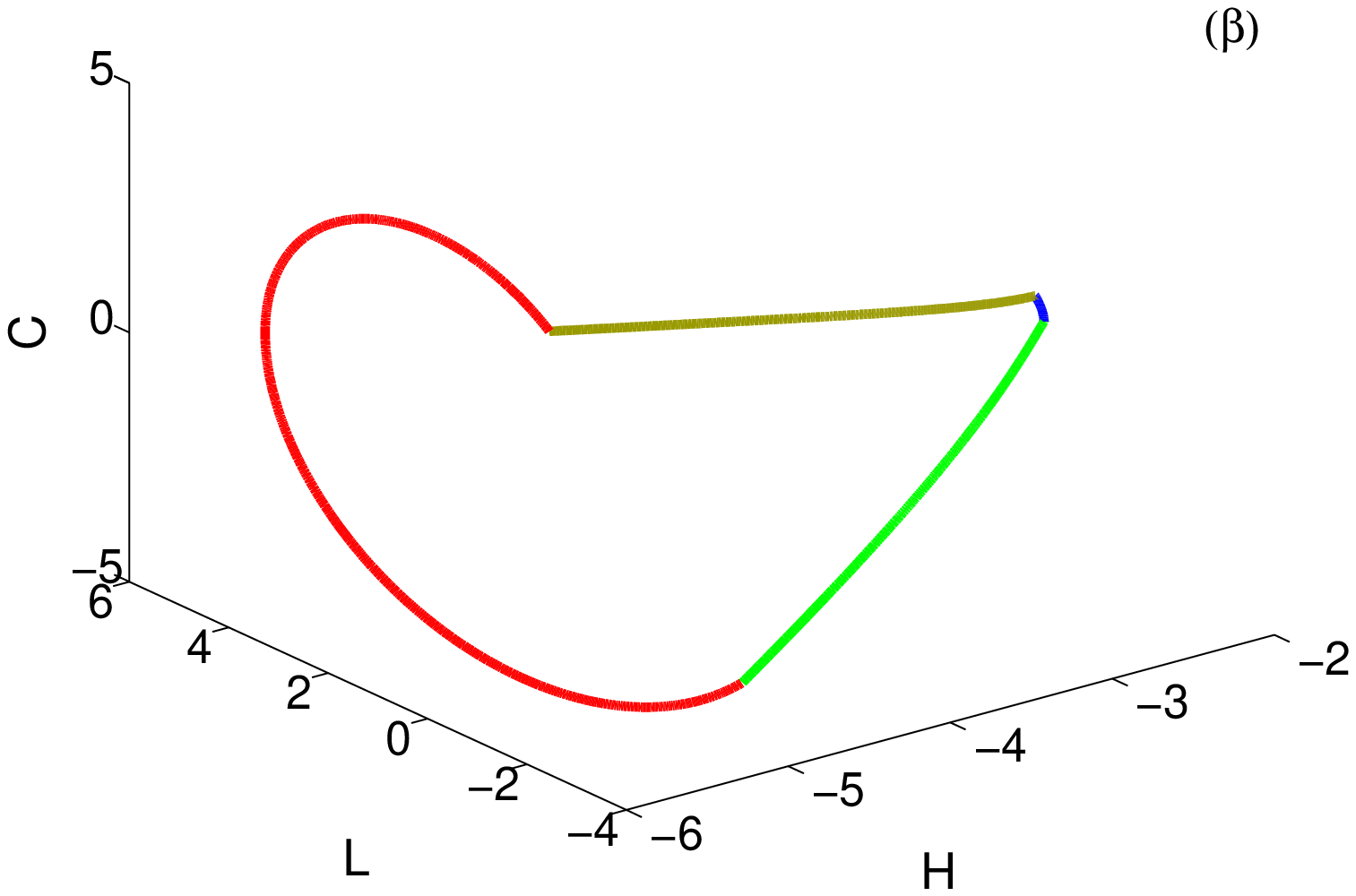}}
\vspace{0.1cm} 
\caption{(Color online)A typical sudden cycle of the refrigerator with additional time allocation on the segments. 
Top($\alpha$): in the ($\Omega,{\cal  S}_E$),~ and ($\Omega,{\cal  S}_{VN}$) 
plane (rectangle), together with the isotherms corresponding to the cold/hot baths 
temperatures, $T_c/T_h$.  Note ${\cal S}_E >{ \cal S}_{VN}$.
The cycle parameters are: $T_c=2.175, T_h=2.9, J=2.5, \omega_c=2.5$, $\omega_h=10.$.
The time allocations are:~$\tau_{c},\tau_{ch}, \tau_h,\tau_{hc}=0.2,0.21,0.44,0.21$, $\kappa_{h,c}=0.36,0.328$.
Bottom($\beta$): the corresponding cycle trajectory in the $H,L,C$ space. 
The longer time allocation on the hot {\em isomagnetic} segment and the larger frequency
$\Omega_h$ allows to complete approximately 3/4 of a period.}
\label{fig:typical}  
\end{figure}

\begin{figure}[tb]
\center{\includegraphics[height=5cm]{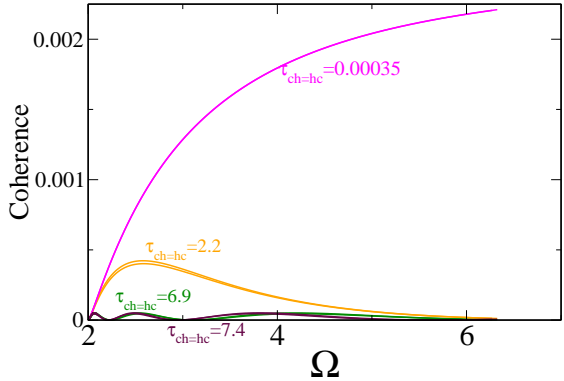}}
\vspace{1.cm}
\caption{(Color online)The coherence $\tilde {\cal C}$, Eq. (\ref{eq:coherence}) as a function 
of $\Omega$ for cycles with different time allocations  on the 
\em adiabats \rm. The cycle parameters correspond to the data of Fig. \ref{fig:typical2}
except the different $\tau_{hc}=\tau_{ch}$. Notice the decrease in coherence when the time allocation on the {\em adiabats} increases.
}
\label{fig:fourcoherence} 
\end{figure}

Sudden cycles can be classified according to the time spent on the different branches.
Fig. \ref{fig:typical2} displays the most extreme case. 
It seems that the two magnetization and demagnetization {\em adiabats}
almost coincide in the entropy frequency plane. Also one recognizes that the cold
isotherm almost touches the only point where the $S_E$ and $ S_{VN}$
meet, showing that further cooling is impossible. 
The  Von Neumann entropy $ S_{VN}=-tr \{{\Op \rho} \ln{\Op \rho}\} $ 
is constant on the \em adiabats \rm since the dynamics on these segments is unitary. 
The bottom of  Fig. \ref{fig:typical2},
shows the trajectories in the $H, L, C$ space indicating that the dependence 
of the different segments forces the trajectory to reside on a plane. 
 
\begin{figure}[tb]
\vspace{1.2cm} 
\center{\includegraphics[height=5cm]{globcyc.eps}}
\center{\includegraphics[height=5cm]{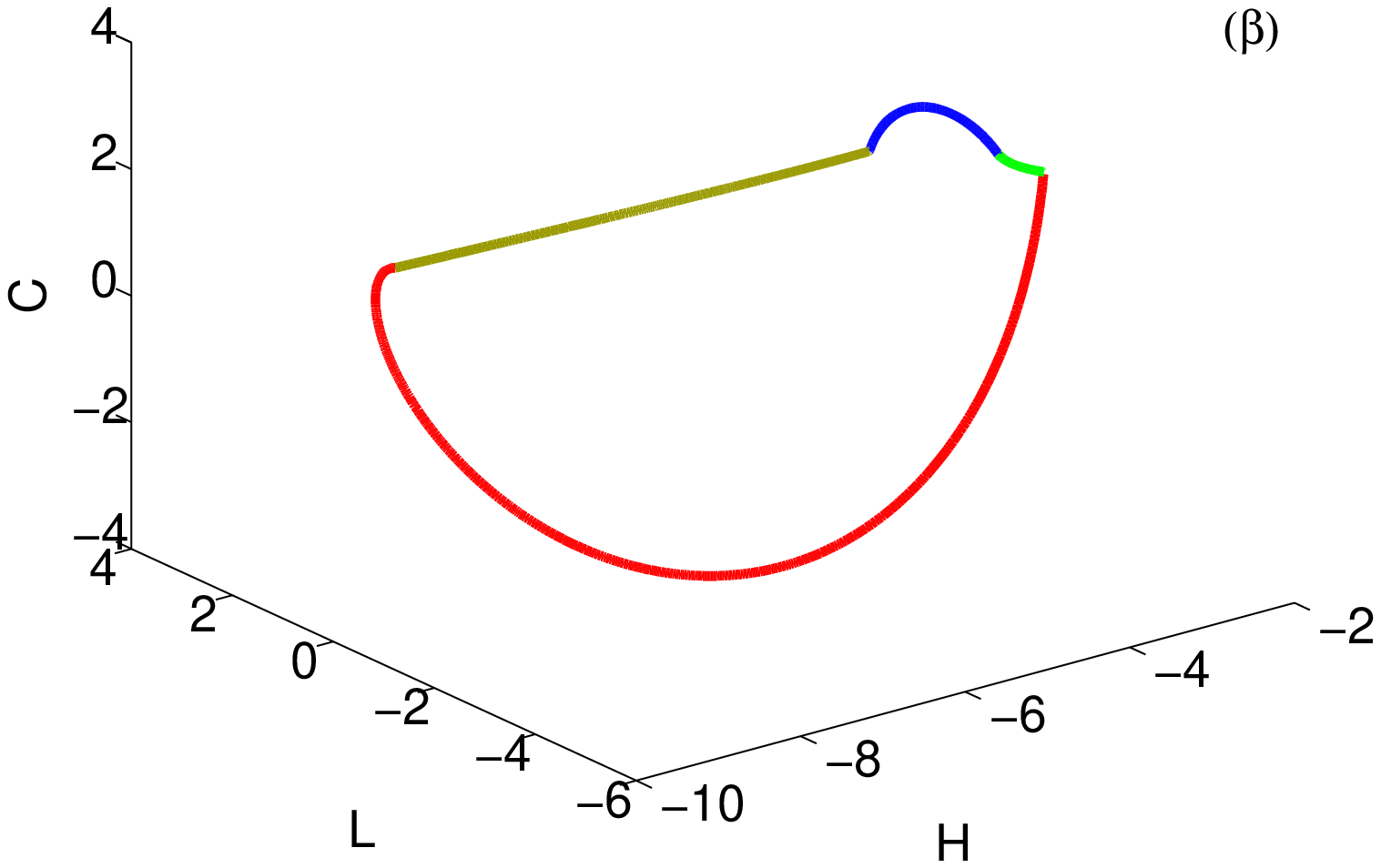}}
\vspace{0.1cm} 
\caption{(Color online)
A  \em sudden \rm cycle with additional time allocated on the {\em isomagnetic} segments. 
The cycle parameters  are:
J=2.5, $T_h=2.9, T_c=2.175, \omega_h=10, \omega_c=2.5$
$k_h \downarrow =3.35, k_c \downarrow = 0.328$
$\tau_h=0.442, \tau_{hc}=0.00744, \tau_c=0.527,\tau_{ch}=0.00824$.
Top($\alpha$):~~In the $(\Omega, {\cal S}_E)$,~$(\Omega, {\cal S}_{VN})$ planes.
Bottom($\beta$):~~The corresponding trajectory in the 
$H,~L,~C$ space. Linear scheduling of $\omega(t)$. 
}
\label{fig:global }            
\end{figure}

Fig. \ref{fig:typical}   shows a second type of sudden cycle with addional time allocation
on the {\em isomagnetic} segments. 
The bottom of Fig. \ref{fig:typical} shows the trajectory of the cycle in 
the $H, L, C$ space. One notices the global property of the cycle,
by realizing that the  \em isomagnetic \rm segments complement each
other (see also the third sudden cycle, Fig. \ref{fig:global }).
The rotation of $L$ and $C$ complete one period split between the hot and cold 
{\em isomagnetic } segments.
In addition, Fig. \ref{fig:typical} shows that further cooling is 
possible, because the end point $B$ of the cold \em isomagnetic \rm segment is 
below the point $E$ where the vertical $BE$ intersects with the cold isotherm.
Point $E$ is the upper-bound on $S_E$ for a cooling cycle.
The maximum heat per cycle that still can be extracted from the cold reservoir
is $T_c({\cal S}_E(E)-{\cal S}_E(B))$.  

One should also notice that in all the sudden cycles, the significant difference 
between $S_E^{max}-S_E^{min}$ and $S_{VN}^{max}-S_{VN}^{min}$,
or generally the difference between  $S_E$ and its corresponding lower
bound $S_{VN}$, all of which indicate the large off diagonal elements of the 
density matrix in the energy representation.  The reason is that while 
$ H,~L ,~C $ are created on the
\em adiabats\rm, the very short time on the hot \em isochores \rm 
is not sufficient to equilibrate the energy of the working fluid.
Generally, the time on the segments is so short, that less than one period is achieved.

\section{Analytical approximations for the cooling power.}
\label{sec:approximations}    

The cooling power is the object of this study.
The amount of heat extracted from the cold bath is defined by the difference in energy 
of the working medium between point $B$ and point $A$ Cf.  Fig. \ref{fig:typical}. 
The cooling power, ${\cal P}_c=Q_c/\tau$, is the heat flow from the cold reservoir into the system divided by $\tau$.
\begin{equation}
{\cal P}_c~~=~~{\cal Q}_c/\tau~~=~~(E_B-E_A)/\tau      .
\label{eq:heatflow} 
\end{equation}
The values of $E_A$ and $E_B$  are calculated from the limit cycle invariant vector $\vec X$ of the 
cycle propagator ${\cal U}_{cyc}$ at these points.

The commutator of the cycles' propagators: 
\begin{equation}
[{\cal U}_{AB}, {\cal U}_c]
\label{eq:heatflow2} 
\end{equation} 
where $ {\cal U}_{AB}= {\cal U}_{hc} {\cal U}_h {\cal U}_{ch}$ supplies an indication 
of the cooling power. When $[{\cal U}_{AB}, {\cal U}_c]=0$, ${\cal Q}_c=0$.
This relation is used to check approximations of the propagators ${\cal U}$.
The sudden limit on the different segments suggest a short time approximation. We use the 
commutator Eq. (\ref{eq:heatflow2}) to check the minimum order of the approximation.

\subsection{Approximations for the \em adiabats \rm}
\label{sec:subapproxadi}

The exact solution for  the propagator on the \em adiabats \rm, 
${\cal U}_{ad}$ for a constant adiabatic parameter $\mu$
has been derived in  Ref. \cite{kosloff10}, Eq. (18):
\begin{eqnarray}
{\cal U}_{ad}~~=~~\frac{\Omega_c}{\Omega_h}
\left(
\begin{array}{ccccc}
\frac {1 + \mu^2 c}{q^2}   & -\frac{\mu s}{q}   &  \frac {\mu(1-c)}{q^2}&0&0   \\
 \frac {\mu s} { q}       & c    &- \frac {s}{q} &0&0  \\
 \frac {\mu (1-c)}{ q^2}  & \frac {s}{q} & \frac {\mu^2+c} {q^2}  &0&0\\
0&0&0&1&0\\
0&0&0&0&\frac{\Omega_h}{\Omega_c}\\
\end{array}
\right) ~~,
\label{eq:calprop}  
\end{eqnarray}   
where  $q= \sqrt{1 +\mu^2}$, $s=sin(q \Theta)$ and $c=cos(q \Theta)$. 
The angle $\Theta$ for the hot-to-cold {\em adiabat} is defined as:
\begin{equation}
\Theta_{hc}=\tau_{adi} (1/K_{hc}
) \left(\arcsin(\omega_c/\Omega_c) -
\arcsin(\omega_h/\Omega_h )\right) \equiv \frac {\tau_{adi}}{K_{hc}} (\Phi_{hc})  
\label{eq:Theta} 
\end{equation}  
A similar expression is obtained for cold-to-hot  angle $\Theta_{ch}$, where:
$K_{hc}=\frac {1}{J} (\omega_c/\Omega_c-\omega_h/\Omega_h)=
-K_{ch}=-\frac {1}{J} (\omega_h/\Omega_h-\omega_c/\Omega_c)$.
The propagator Eq. (\ref{eq:calprop}) is proportional to the 
compression ratio: ${\cal C}=\frac{\Omega_c}{\Omega_h}$ 
except for the last term which corresponds to  the identity.

Expanding the terms composing  ${\cal U}_{hc}$ for short time 
$(\tau_{ch}= \tau_{hc} \equiv \tau_{adi}  \ll 1)$, one obtains
$|\mu|=q$. This result follows from the
definitions of $\mu$ and $q$:
~$\mu_{hc}=-\mu_{ch}=\frac {K_{hc}}{\tau_{adi}}=-\frac {K_{ch}}{\tau_{adi}}$,
where the relation  between $\mu$ and $\tau_{adi}$ is obtained 
from \cite{kosloff10}. Since $\tau_{adi} \ll 1$, from the definition of $\mu$ one obtains that
$|\mu| \gg 1 $.
Further:~$q=\sqrt{1+\mu^2}$. Therefore  $|\mu|=q$.
Under these conditions  ${\cal U}_{hc}$, Eq. (\ref{eq:calprop}), is approximated as:
\begin{eqnarray} 
{\cal U}_{hc}^{(appr)}  \approx~\frac{\Omega_c}{\Omega_h}
\left(
\begin{array}{ccccc}
 c  &-  s  & 0& 0  & 0 \\
 s     &c  & 0 & 0 & 0  \\
0  & 0  &1  & 0 & 0 \\ 
 0  &  0 & 0 & 1  & 0 \\
0  &  0   & 0  & 0 & \frac{\Omega_h}{\Omega_c}  \\
\end{array}
 \right)
\label{eq:propadhc}      
\end{eqnarray} 
The explicit expression for the argument $q \Theta $ becomes:
$q \Theta = |\mu| \Theta = |K|/\tau_{adi}~ \Theta=|K|/\tau_{adi} 
\frac {\tau_{adi}}{|K|} (\Phi_{hc})=  (\Phi_{hc}) $.
A similar expression is obtained for ${\cal U}_{ch}^{(a)} $.
Notice that in  Eq. (\ref{eq:propadhc}), only  the vector components 
$\Op H$ and $\Op L$ exchange their values.

\subsubsection{Further classification of sudden approximations on the \em adiabats \rm.}
\label{sec:partitionapprox}   

A further classification of the propagators on the
\em adiabats \rm is defined according to the values of the operating parameters
$\omega_c, \omega_h $ and $J$ is based on additional 
simplification of either  Eq. (\ref{eq:calprop}) or Eq. (\ref{eq:propadhc}).
 
\begin{enumerate}

\item

When $\omega_c \ll J$, $ \omega_h \gg J$ and also
$\tau_{adi} \ll \frac {2 \pi}{\Omega(t)}$ in the range of t, then the propagator
${\cal U}_{hc}^{(appr)}$ will simplify to the form: 
\begin{eqnarray}
 {\cal U}_{hc}^{(1)}=~\approx~\frac{\Omega_c}{\Omega_h}
\left(
\begin{array}{ccccc}
0  & 1 & 0& 0  & 0 \\
-1     &  0  & 0 & 0 & 0  \\
0  & 0  &1  & 0 & 0 \\ 
 0  &  0 & 0 &  1  & 0 \\
0  &  0   & 0  & 0 & \frac{\Omega_h}{\Omega_c}  \\
\end{array}
 \right)
\label{eq:propadihc1}    
\end{eqnarray}

\item
When $\omega_c \sim J$, $ \omega_h \gg J$
and also $\tau_{adi} \ll \frac {2 \pi}{\Omega(t)}$. 
In this case ${\cal U}_{hc}^{(appr)}$  will have the form :
\begin{eqnarray}
 {\cal U}_{hc}^{(2)}~\approx~\frac{\Omega_c}{\Omega_h}
\left(
\begin{array}{ccccc}
\frac {1}{\sqrt{2}}   & \frac {1}{ \sqrt{2}}   
 & 0& 0  & 0 \\
- \frac {1}{\sqrt{2}}  & \frac {1}{ \sqrt{2}} 
   & 0 & 0 & 0  \\
0  & 0  & 1 & 0 & 0 \\ 
 0  &  0 & 0 &  1  & 0 \\
0  &  0   & 0  & 0 & \frac{\Omega_h}{\Omega_c}  \\
\end{array}
\right)
\label{eq:propadihc2}  
\end{eqnarray} 

\item
case(a)

When $\omega_c ,~ \omega_h \gg J$
and also $\tau_{adi} \ll \frac {2 \pi}{\Omega(t)}$. 
then ${\cal U}_{hc}^{(appr)}$ becomes:
\begin{eqnarray} 
 {\cal U}_{hc}^{(3a)}~\approx~\frac{\Omega_c}{\Omega_h} 
\left(
\begin{array}{ccccc}
1  & 0   
 & 0& 0  & 0 \\
0  & 1
   & 0 & 0 & 0  \\
0  & 0  & 1 & 0 & 0 \\ 
 0  &  0 & 0 & 1  & 0 \\
0  &  0   & 0  & 0 & \frac{\Omega_h}{\Omega_c}  \\
\end{array}
\right)
\label{eq:propadihc3a}    
\end{eqnarray} 

Case(b)
 
When $\omega_c ,~ \omega_h \gg J$  and also $ \tau_{adi} \approx 1$. \\
>From the condition  $\omega_c ,~ \omega_h \gg J$ the ratio $\frac {\omega}{\Omega}$ becomes:
\begin{equation}
\frac {\omega}{\Omega}=1-(1/2) (\frac {J}{\omega})^2.
\label{eq:omegaratio}  
\end{equation} 
>From the definition of $\mu$ in Eq (\ref{eq:calprop}), $K_{adi}$ after Eq. (\ref{eq:Theta})  
and condition (b), follows $K_{adi} <1$, therefore $\mu <1 $.
In addition from the definition of $q$ follows  $q=1$. 
Therefore   Eq. (\ref{eq:calprop}) simplifies by neglecting $\mu^2$ to the following propagator:
\begin{eqnarray}
{\cal U}_{hc}^{(3b)}~~=~~\frac{\Omega_c}{\Omega_h}
\left(
\begin{array}{ccccc}
 1    & -\mu s   &  \mu(1-c)&0&0   \\
 \mu s  & c    &- s &0&0  \\
 \mu (1-c)  & s & c   &0&0\\
0&0&0&1&0\\
0&0&0&0&\frac{\Omega_h}{\Omega_c}\\
\end{array}
\right) ~~,
\label{eq:calprop3b}    
\end{eqnarray}   
In ref. \cite{kosloff10}, when  expanding $H,L,C$ in Eq. (\ref{eq:calprop})
to first order in $\mu$, a propagator similar to Eq. (\ref{eq:calprop3b}) was 
obtained.
  
>From $q=1$, Eqs. (\ref{eq:omegaratio}) and (\ref{eq:Theta}), the argument 
$q \Theta$ of the trigonometric functions becomes
\begin{equation}
q \Theta_{hc}= 1 \tau_{adi} 
\frac {(J^2/2)\left(1/\omega_h^2-1/\omega_c^2\right)} 
{(J/2)\left(1/\omega_h^2-1/\omega_c^2\right)}=\tau_{adi} J 
\label{eq:qTheta} 
\end{equation} 
In this case the argument $q \Theta~=~\Theta~$ of the trigonometric functions is large
therefore it cannot be approximated to first (or second)  order.  

\end{enumerate}

\subsection{Approximations for the \em isomagnetic segments \rm.}
\label{sec:isomagnetapprox}

For the \em isomagnetic \rm segments the propagators ${\cal U}_c$ and 
${\cal U}_h$ are approximated as follows.

\begin{enumerate}

\item  {
 
First we assume that the time allocation on both \em isomagnetic \rm  segments
is short enough, so that
in Eq.  (\ref{eq:propiso}),~  $ \sin(\Omega \tau) =0$, and $\cos(\Omega \tau)=1$.
In addition, $\gamma=0$. Then Eq.  (\ref{eq:propiso})  
simplifies to:
\begin{eqnarray} 
{\cal U}_{i}^{1}~\approx~
\left(
\begin{array}{ccccc}
e^{(-\Gamma_{c/h} \tau)}   & 0 & 0& 0  & {E}_{eq}
(1-e^{(-\Gamma_{c/h} \tau)}) \\
0  & e^{(-\Gamma_{c/h} \tau)}       & 0  & 0 & 0  \\
0  & 0   &  e^{(-\Gamma_{c/h} \tau)}   & 0 & 0 \\ 
\frac {1}{\Omega}(E_{eq}(e^{-\Gamma_{c/h}\tau}-e^{-2\Gamma_{c/h}\tau}))   &  0 & 0 & e^{-2\Gamma_{c/h} \tau}  & \frac{E_{eq}^{2}}{\Omega}(e^{-\Gamma_{c/h} \tau}-1) \\
0  &  0   & 0  & 0 & 1  \\
\end{array}
\right)
\label{eq:propisoapr}  
\end{eqnarray}   

}

\item  {
On the \em isomagnetic \rm segments we assume short time on the \em hot isomagnetic \rm segment, 
leading to a first order approximation in time.
The result is an  approximate propagator for the  \em hot isomagneticic \rm segment: 
\begin{eqnarray} 
 {\cal U}_{hot}^{2}~\approx~
\left(
\begin{array}{ccccc}
{1-\Gamma_h \tau_h}    & 0 & 0& 0  & (E_h^{eq})(\Gamma_h \tau_h)  \\
0  & {1-\Gamma_h \tau_h} & -\Omega_h \tau_h  & 0 & 0  \\
0  &\Omega_h \tau_h   &  {1-\Gamma_h \tau_h}   & 0 & 0 \\ 
(\frac {1}{\Omega_h}) (E_h^{eq})(\Gamma_h \tau_h)  &  0 & 0 & 
1-2 \Gamma_h \tau_h  & \frac{E_h^{eq}{2}}{\Omega_h}(\Gamma_h \tau_h) \\
0  &  0   & 0  & 0 & 1  \\
\end{array}
\right)
\label{eq:prophotisoapr}   
\end{eqnarray}   
A first order approximation also on the cold segment leads to the 
commutation of the segment propagators Eq. (\ref{eq:heatflow2}) 
to vanish $[{\cal U}_{AB}, {\cal U}_c]=0$. 
Therefore for the cold segment, the unapproximated 
Eq. (\ref{eq:propiso}) is employed. 

Another possibility is that the hot and cold \em isomagnetic \rm segments are swapped, and the time
on the \em cold  isomagnetic \rm segment is short, while the \em hot isomagnetic \rm
is not  approximated.

}

\item {

In this case both of the \em isomagnetic \rm segments are approximated to the second order in time
leading to:
\begin{eqnarray}
\nonumber 
 {\cal U}_{c/h}^{3}~\approx~
\end{eqnarray}
\begin{eqnarray}  
\left(
\begin{array}{ccccc}
1-G+\frac {G^2}{2} & 0 & 0& 0  & E^{eq}(G- \frac {G^2}{2} ) \\
0  & {1-G+\frac {(\Gamma^2-\Omega^2)\tau^2}{2} } & -\Omega \tau+\Gamma 
\Omega \tau^2 & 0 & 0  \\
0  &\Omega \tau-\Gamma \Omega \tau^2 & 1-G +\frac {(\Gamma^2-
\Omega^2)\tau^2}{2} & 0 & 0 \\ 
\frac {E^{eq}}{\Omega}(G-\frac {3G^2}{2})&  0 & 0 & 
1-2 G +2G^2 & \frac{E^{eq 2}}{\Omega}(G 
 -\frac{G^2}{2}) \\
0  &  0   & 0  & 0 & 1  \\
\end{array}
\right)
\label{eq:propisosecapr}     
\end{eqnarray}   
}
and $G=\Gamma \tau$.
\end{enumerate}

\subsection{The approximate cooling power} 
\label{sec: synthesis}

The heat removed from the cold bath at each period is calculated from
the global propagator, Eq. (\ref{eq:globalprop}). The approximations of the segment propagators 
are employed, Eqs. (\ref{eq:propadihc1}) to (\ref{eq:propisosecapr}), to obtain a global propagator. 
The next step is to evaluate the invariant vector of the global propagator with eigenvalue one.
Obtaining such a vector is an internal verification on the validity of the approximation.
The energy $E$ component of the eigenvector 
at two points $B$ and $A$ in Fig.  \ref{fig:typical} lead to ${\cal Q}_c=E_B-E_A$.

\begin{enumerate}

\item  {

In the first combination  the \em adiabats \rm  will be approximated
by Eq. (\ref{eq:propadihc1}), while the  \em hot isomagnetic \rm segment
will be approximated by Eq. (\ref{eq:prophotisoapr}), and the 
 \em cold isomagnetic \rm by Eq. (\ref{eq:propiso}).
Denoting $e^{-(\Gamma_c \tau_c)}= \alpha $ and 
$\cos(\Omega_c \tau_c)=cc,  \sin(\Omega_c \tau_c)=ss  $,
the  heat removed from the cold bath in this
approximation, $ {\cal Q}_c^{appr1} $ becomes:
 
\begin{equation}
{\cal Q}_c^{appr1}  ~\approx~ - \frac{ \alpha  \tau_h^2(\Omega_c ss
 E_h^{eq} \Gamma_h - \Omega_h^2 E_c^{eq}  cc
+E_c^{eq} \alpha \Omega_h^2)}{
(\alpha^2-2 \alpha cc +1)}+\frac{(2 \alpha \tau_h^2 \Gamma_h^2 E_c^{eq}(\alpha-cc))}{(\alpha-1)}  
\label{eq:Qcsecappr1}           
\end{equation}  

The second term in Eq. (\ref{eq:Qcsecappr1}) is typically two order of 
magnitude smaller than the first term, therefore we neglect it. The final  
approximation in this case becomes:

\begin{equation}
{\cal Q}_c^{appr1b}  ~\approx~ - \frac{ \alpha  \tau_h^2(\Omega_c ss
 E_h^{eq} \Gamma_h - \Omega_h^2 E_c^{eq}  cc
+E_c^{eq} \alpha \Omega_h^2)}{(\alpha^2-2 \alpha cc +1)}
\label{eq:Qcsecappr1b}           
\end{equation}

The approximations of Eq.  (\ref{eq:Qcsecappr1}) and (\ref{eq:Qcsecappr1b}) 
correspond to the data of  Fig. \ref{fig:typical2}.
The approximation was compared to the numerical  calculations
in the range of parameters corresponding  to Fig. \ref{fig:typical2},  
in a neighborhood of $\tau_c$, and  $T_c$. A good agreement was obtained up to 
a constant numerical factor.

The heat removed ${\cal Q}_c$ in Eq. (\ref{eq:Qcsecappr1b}) can change sign meaning that for certain values of parameters the refrigeration stops. Fig. \ref{fig:signswitch} shows the alternating cooling  as a function of $\tau_c$.
The switching points of approximation Eq. (\ref{eq:Qcsecappr1b}) are conjectured 
at the points where the functions $\exp(\Gamma_c \tau_c)$ 
and $\cos(\Omega_c \tau_c)$ cut each other, (Bottom of Fig. \ref{fig:signswitch}). 
As can be seen these points form a good approximation.
    
Fig. \ref{fig:seepisland} is a map of showing regions in parameter space where
refrigeration takes place and regions where there is no refrigeration. 
In this map $\tau_c=\tau_h$, which is a border case of 
the approximation, because the approximation in Eq. (\ref{eq:Qcsecappr1b}) 
assumes $\tau_c/\tau_h \gg 1  $.

}
   
\item 
{
In the second case the \em adiabats \rm  are approximated
by Eq. (\ref{eq:propadihc2}). On both \em isomagnets \rm  
a second order approximation is used, Eq. (\ref{eq:propisosecapr}),  leading to:
\begin{equation}
{\cal Q}_c^{appr2}~\approx~\frac{( E_c^{eq} - \frac {\Omega_c E_h^{eq}}{\Omega_h})
(\Gamma_c \tau_c-\frac{1}{2}(\Gamma_c \tau_c)^2)(\Gamma_h \tau_h-0.5(\Gamma_h \tau_h)^2)}{ (\Gamma_c \tau_c+\Gamma_h \tau_h)-\frac{1}{2}(\Gamma_c \tau_c+\Gamma_h \tau_h)^2}
\label{eq:Qcsecappr2}     
\end{equation}
Fig. \ref{fig:threehot } shows cycles
corresponding to the  conditions of Eq.  (\ref{eq:Qcsecappr2}),
with the additional condition: $\omega_h \tau_h=constant=6.252$.  
Comparison of the approximation of Eq. (\ref{eq:Qcsecappr2}) to numerical values of ${\cal Q}_c$
show good agreement with deviations up to $\sim 20 \%$.
Section \ref{sec:coolinginf} addresses a large subfamily of
cycles corresponding to Eq. (\ref{eq:Qcsecappr2}), with the additional condition  of
$\omega_h \tau_h=constant$.
see Fig. \ref{fig:threeappr2}.
\begin{figure}[tb]
\vspace{2.2cm} 
\center{\includegraphics[height=5cm]{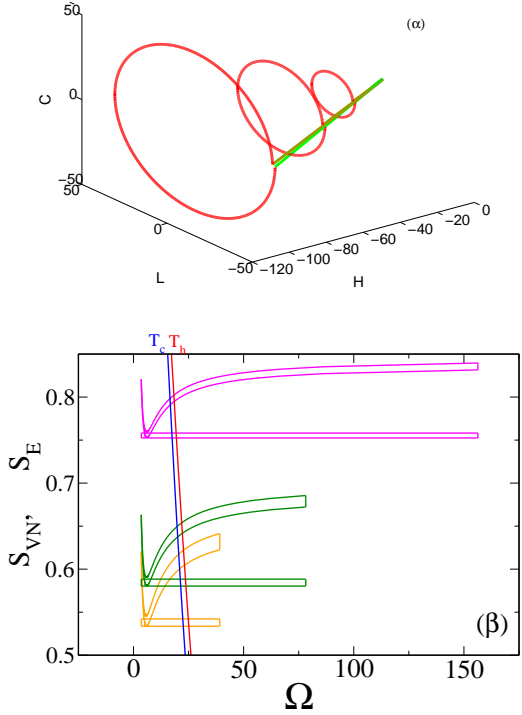}}
\center{\includegraphics[height=5cm]{cyc452453454.eps}}
\vspace{0.1cm} 
\caption{(Color online)
Top($\alpha$):~~Cycle trajectories in the $H,L,C$ space which complete approximately one revolution
on the hot {\em isomagnetic } segment. $\omega_h \tau_h=6.252$.
 The data for the three
cycles are: $J=2.5, T_h=10, T_c=9., \omega_c=2.5$
$k_h \downarrow =0.36, k_c \downarrow = 0.328$
$\tau_{hc}=\tau_{ch}=0.065625,\tau_{c}=0.008375$, 
>From large, to small cycles:
$\omega_h=156.3,78.15,39.075$ and therefore $\tau_h=0.04,0.08,0.16$
Bottom($\beta$):~~The same cycles in the $(\Omega, {\cal S}_E), (\Omega, {\cal S}_{VN})$, planes. 
}
\label{fig:threehot }            
\end{figure}   
}
\item

{

In the third approximation we will distinguish between two cases.
 
Case A;

First we combine Eq. (\ref{eq:propadihc3a})
for the \em adiabats \rm  and Eq. (\ref{eq:propisoapr}) for
the \em isomagnets\rm. This case becomes equivalent (for $(\Gamma_c \tau_c+\Gamma_h \tau_h) < 2$)  
to the frictionless cycles studied before \cite{kosloff10}.
This means that there is no coupling between the energy $E$ and the other variables $L$ and $C$.
As a result for short times  a bang-bang type solution is optimal \cite{feldmann00}.
Expanding the exponents to second order one gets the following expression:
\begin{equation}
{\cal Q}_c^{appr3a}~\approx~E_h \frac{( \frac{E_c^{eq}}{E_h^{eq}} - \frac {\Omega_c }{\Omega_h})
(\Gamma_c \tau_c \Gamma_h \tau_h)}{ (\Gamma_c \tau_c+\Gamma_h \tau_h) 
-(\Gamma_c \tau_c+\Gamma_h \tau_h)^2  }
\label{eq:Qcsecappr3a}     
\end{equation}  
It it can be shown using Eq. (\ref{eq:basicineq}) that ${\cal Q}_c^{appr3}~\geq~0$
for $(\Gamma_c \tau_c+\Gamma_h \tau_h) < 2$. 
} 

Case B;

Combining Eq. (\ref{eq:calprop3b})      
for the \em adiabats \rm  and Eq. (\ref{eq:propisoapr}) for
the \em isomagnets \rm leads to case B. Computing the eigenvalues of the
corresponding global propagator, the variables $H_A$ and $H_B$ separates beautifully
from the other operator expectation values. The result: 
\begin{equation}
{\cal Q}_c^{appr3b}~\approx~E_h \frac{( \frac{E_c^{eq}}{E_h^{eq}} - \frac {\Omega_c }{\Omega_h})
(\Gamma_c \tau_c \Gamma_h \tau_h) -2 \frac{E_c^{eq}}{E_h^{eq}} (1-A_c)A_h \mu^2(1-c)    }
{ 1-A_h A_c(1+2 \mu^2(1-c))  }
\label{eq:Qcsecappr3b}      
\end{equation} 
where $c=\cos(J \tau_{adi})$ and $A_{h/c}=\exp(-\Gamma_{h/c} \tau_{h/c})$.
Fig. \ref{fig:ansignchange} plots Eq. (\ref{eq:Qcsecappr3b}) as a function
of $\tau_{adi}$ with some internal scaling. The approximation
demonstrates the sign changes of the heatflow.
The cycles of Fig.  \ref{fig:concconv}   correspond to 
Eq. (\ref{eq:Qcsecappr3b}).
In Section \ref{sec:discontinuity} we will further elaborate on the
properties of those cycles and their approximations.
\end{enumerate}

\subsection{The discontinuous character of the \em sudden \rm cycle families}
\label{sec:discontinuity}      

\begin{figure}[tb]
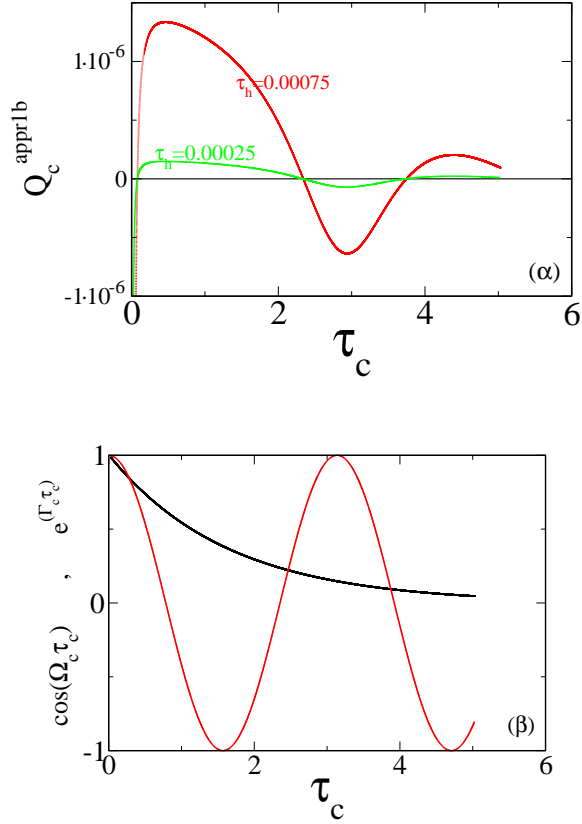

\vspace{1.2cm} 
\center{\includegraphics[height=5cm]{Qcappr1b.eps}}
\vspace{0.5cm} 
\center{\includegraphics[height=5cm]{cospexp.eps}}
\vspace{1.1cm}  
\caption{(Color online) 
Top($\alpha$): Eq. (\ref{eq:Qcsecappr1b}) is presented when changing the values of $\tau_c$,
for two different $\tau_h$ values, as denoted on the figure:
 $\tau_h=0.00075$  red line ,  $\tau_h=0.00025$   green line.
Bottom($\beta$): The functions $\exp(\gamma_c \tau_c)$ and $\cos(\Omega_c \tau_c)$
are shown as functions of $\tau_c$. Their crossing points compare well to the exact points(Top figure).
The other parameters are $J=2, \tau_{ch}=\tau_{hc}=0.00035, \omega_c=0.1,
 \omega_h=6, T_h=15, T_c=14 $, as in Fig.  \ref{fig:typical2}.
} 
\label{fig:signswitch}        
\end{figure} 

\begin{figure}[tb]
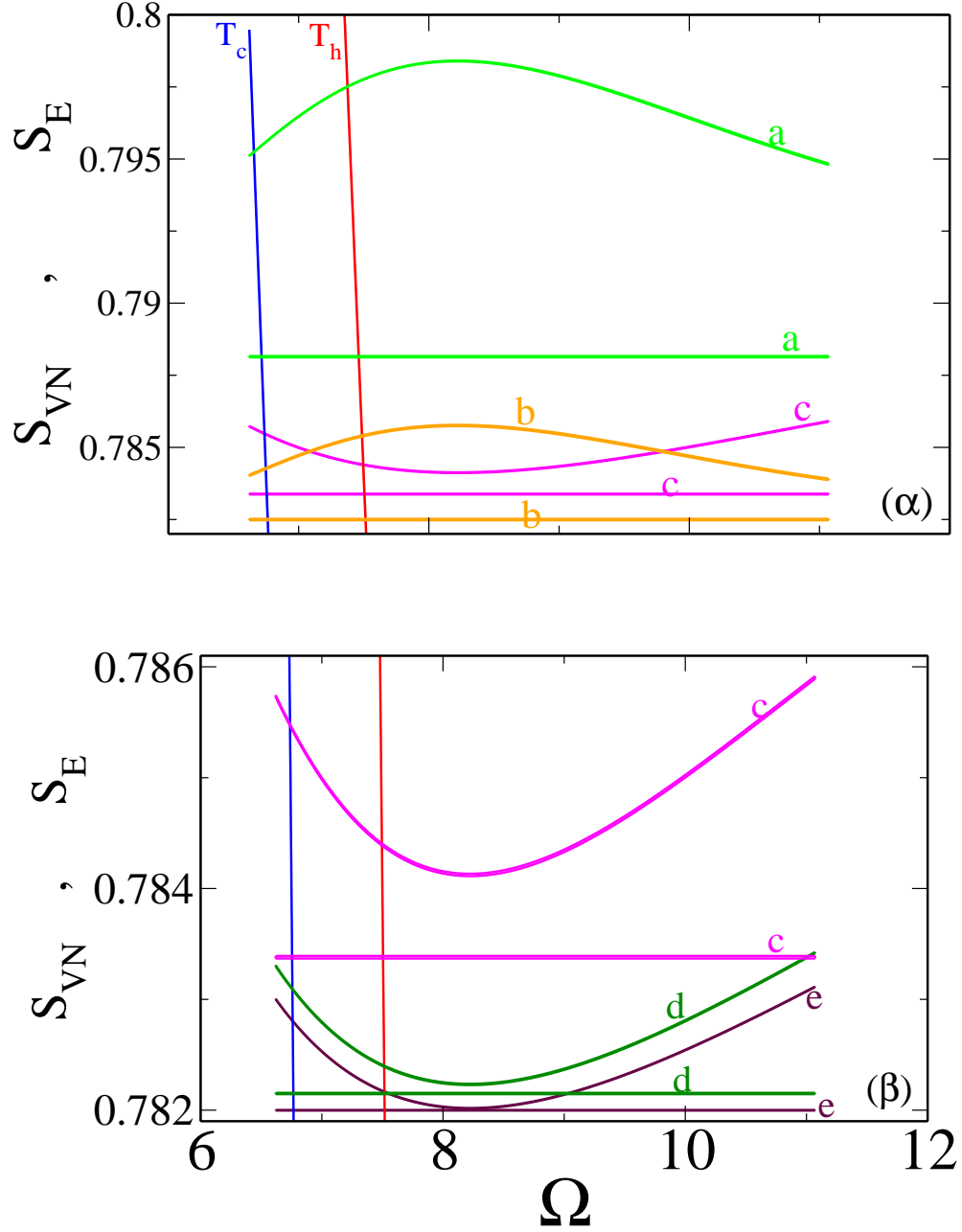

\vspace{1.2cm} 
\center{\includegraphics[height=8cm]{twoconv1conc.eps}}
\vspace{0.6cm} 
\center{\includegraphics[height=8cm]{threelowt.eps}}
\vspace{0.1cm}  
\caption{(Color online) 
Top($\alpha$): ${\cal S}_E$ and ${\cal S}_{VN}$ as a function of $\Omega$ cycles
for three different sets of times~, $\tau_{h} ,\tau_{ch}=\tau_{hc},\tau_{c}=$~are:
~0.000405,~0.4194,~0.029538~green (a),~0.00045,~0.466,~0.03282~orange (b),
~0.00036,~0.3728,~0.026256~ non refrigeration cycles,
~0.000225,~0.233,~0.01641~magenta (c) cycles.
~~Bottom($\beta$): Continuation of Top by lowering the cycle time proportionally.
 $\tau_{h} ,\tau_{ch}=\tau_{hc},\tau_{c}=$~are: magenta (c) as on Top,
~0.0001125,~0.1165,~0.008205~green (d), 
~0.00005625,~0.05825,~0.0041025~maroon (e)
The parameters for all the cycles are:
$J=1.25, T_h=4, T_c=3.6, \omega_h=11, \omega_c=6.5 $,
$ \kappa_h^{\downarrow}=0.36, \kappa_c^{\downarrow}=0.0656$, see 
Eq. (\ref{eq:Qcsecappr3b})  } 
\label{fig:concconv}        
\end{figure}

\begin{figure}[tb]
\center{\includegraphics[height=5cm]{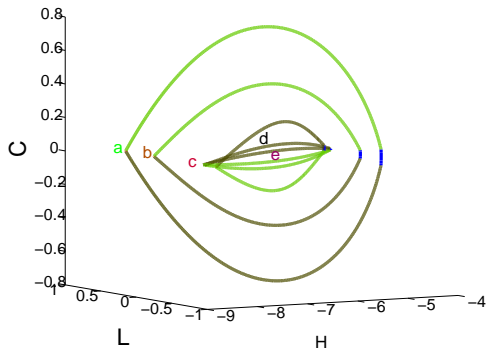}}
\vspace{1.cm}
\caption{(Color online)The trajectories in the space $H,~ L,~C$  of the cycles of 
Fig. \ref{fig:concconv} where the cycles are indicated by letters corresponding 
to the cycles of Fig. \ref{fig:concconv}. }
\label{fig:threelowt} 
\end{figure}

An example of the discontinuous behavior is shown in Fig. \ref{fig:concconv}   corresponding to 
Eq. (\ref{eq:Qcsecappr3b}). When the time allocation 
on the \em adiabats \rm is increased the cycles changed from  a concave shape to  a convex shape.
Two of these cycles are shown in Fig. \ref{fig:concconv}. Additional reduction in  cycle
times leads to ${\cal Q}_c < 0$, which means that the cycles cease to be refrigerators. Then by further reducing
the allocated times, the cycles suddenly transformed into a concave shape.
In addition ${\cal Q}_c > 0$, refrigerator cycles again.   
Fig.  \ref{fig:concconv}  shows an example for the stated behavior
of the \em sudden \rm cycles; there are families of cycles
with a small $\delta$ neighborhood, beyond which a discontinuity emerges. Changing parameters can lead to
another  small  neighborhood. In Fig.  \ref{fig:concconv}  the times on the segments were changed proportionally,
decreasing  the overall cycle time, so that $\tau_{cycle} \rightarrow 0$. 
Fig. \ref{fig:threelowt} presents the cycles of  Fig. \ref{fig:concconv} 
in the  $H,~ L,~C$ space.

\begin{figure}[tb]
\vspace{1.cm}
\center{\includegraphics[height=5cm]{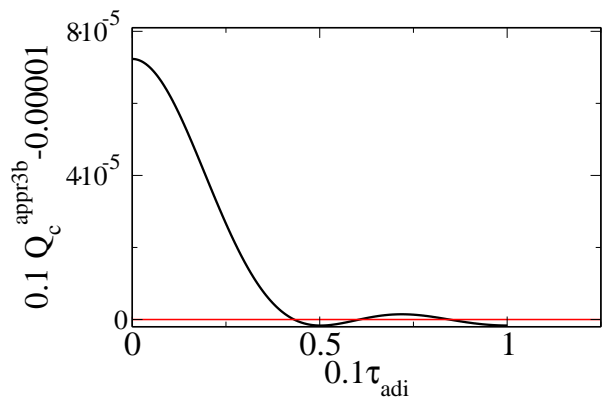}}
\vspace{1.cm}
\caption{(Color online)
$Q_c$ as a function of $\tau_{adi}$ with some internal scaling
as denoted on the figure
in order to achieve the exact results by the approximation of
 Eq. (\ref{eq:Qcsecappr3b}), which  
approximate the cycles of Fig. \ref{fig:concconv},
with the same parameters. The fineness of the sign change is obvious.  }
\label{fig:ansignchange}  
\end{figure} 

\begin{figure}[tb]
\vspace{0.66cm}
\hspace{10.cm} 
\center{\includegraphics[height=7cm]{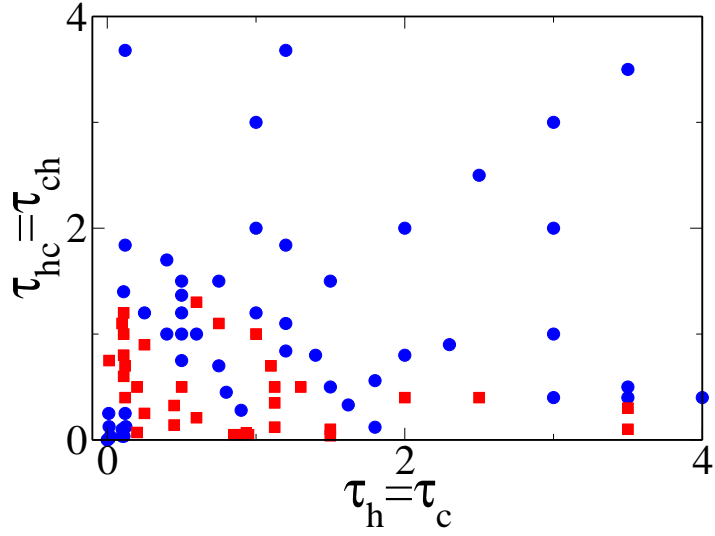}}
\vspace{1.cm}  
\caption{(Color online)A map of parameter regions of time allocation leading to 
refrigeration in blue cycles and non refrigeration regions in red squares. 
The cycle of Fig. 1 is located in the blue island in the lower left corner.
The general parameters are:
 $J=2.,~T_c=14,~T_h=15,
\omega_c=0.1, \omega_h=6$, $\kappa_h^{\downarrow}=0.36, 
\kappa_c^{\downarrow}=0.328$. The parameters used fit case 1: 
Eq. (\ref{eq:Qcsecappr1b}).   
}
\label{fig:seepisland}            
\end{figure}

\section{The Coefficient Of Performance (COP) and the Entropy Generation
(${\cal S}^u$)  } 
\label{COPSU}

The coefficient of performance ($\bf COP$) \rm is defined as the heat extracted divided
by the work input:
\begin{equation}
{\bf COP} \rm ~\approx~   \frac{{\cal Q}_c}{{\cal W}^{on}}
\label{apprCOP}   
\end{equation}  

The entropy generation (${\cal S}^u$) for a cyclic process is generated in the baths:
\begin{equation}
{\cal S}^u ~\approx~-\left( \frac{{\cal Q}_c}{T_c}~+~\frac{{\cal Q}_h}{T_h} \right)
\label{apprentr1}    
\end{equation}   
An explicit approximations for $\bf COP \rm$ and ${\cal S}^u$ for the case 
of Eq. (\ref{eq:Qcsecappr1b}) is now evaluated.
This requires to  compute the work input, ${\cal W}^{on}$.
Using the notation of Fig. \ref{fig:typical}:
\begin{equation}
{\cal W}^{on} ~\approx~ (E_C-E_D)~-~{\cal Q}_c^{appr1b}
\label{apprworkoncyc}   
\end{equation} 
Similarly to Eq. (\ref{eq:heatflow}), the values of $E_C$ and $ E_D$ are
evaluated from the limit cycle invariant vector $\vec X$ of the cycle
propagator at points $C$ and $D$.

Denoting again $e^{-(\Gamma_c \tau_c)}= \alpha $ and 
$\cos(\Omega_c \tau_c)=cc,  \sin(\Omega_c \tau_c)=ss  $,
the work done on the cycle (${\cal W}^{on}$)  becomes:
\begin{equation}
{\cal W}^{on}\approx \frac{\tau_h E_h^{eq} \Gamma_h  
(1-\alpha) (\alpha^2-2 \alpha cc +1)  
+\tau_h^2 \left(E_h^{eq} \Gamma_h^2 \alpha^2 ss^2(2 \alpha-1)-
\frac{\Omega_h^2}{\Omega_c}ss \Gamma_h E_c^{eq} \alpha (1-\alpha)
 \right)
}{
(\alpha-1)(\alpha^2-2 \alpha cc +1)  +
\Omega_h^2 \tau_h^2 \alpha^2 (\alpha-cc) 
-\alpha^2 \Gamma_h \tau_h (\alpha - 2 \alpha \Gamma_h \tau_h  
+ 2 \Gamma_h \tau_h cc)}-{\cal Q}_c  
\label{eq:Workonappr1}   
\end{equation}  
when  terms with third and larger orders of $\tau_h$ are neglected.
Noticing that the lowest order of $\tau_h$ in the expressions for
${\cal Q}_c$ were second order. The invested work becomes:
\begin{equation}
{\cal W}^{on} ~\approx~  - \tau_h E_h^{eq} \Gamma_h  
\label{eq:workfirstor}    
\end{equation}
Eq. (\ref{eq:workfirstor}) shows that the invested work is on the
cold $\rightarrow$ hot \em adiabat \rm and is dissipated almost exclusivly
on the hot {\em isomagnetic} segment.

The $\bf COP \rm$ of the cycle is approximated as:
\begin{equation}
\bf COP \rm ~\approx~\tau_h  \frac{ \alpha (\Omega_c ss
 E_h^{eq} \Gamma_h - \Omega_h^2 E_c^{eq}  cc
+E_c^{eq} \alpha \Omega_h^2) }{ E_h^{eq} \Gamma_h
(\alpha^2-2 \alpha cc +1)} 
\label{eq:COPAPPR12}   
\end{equation}

\begin{figure}[tb]
\vspace{0.3cm} 
\center{\includegraphics[height=10cm]{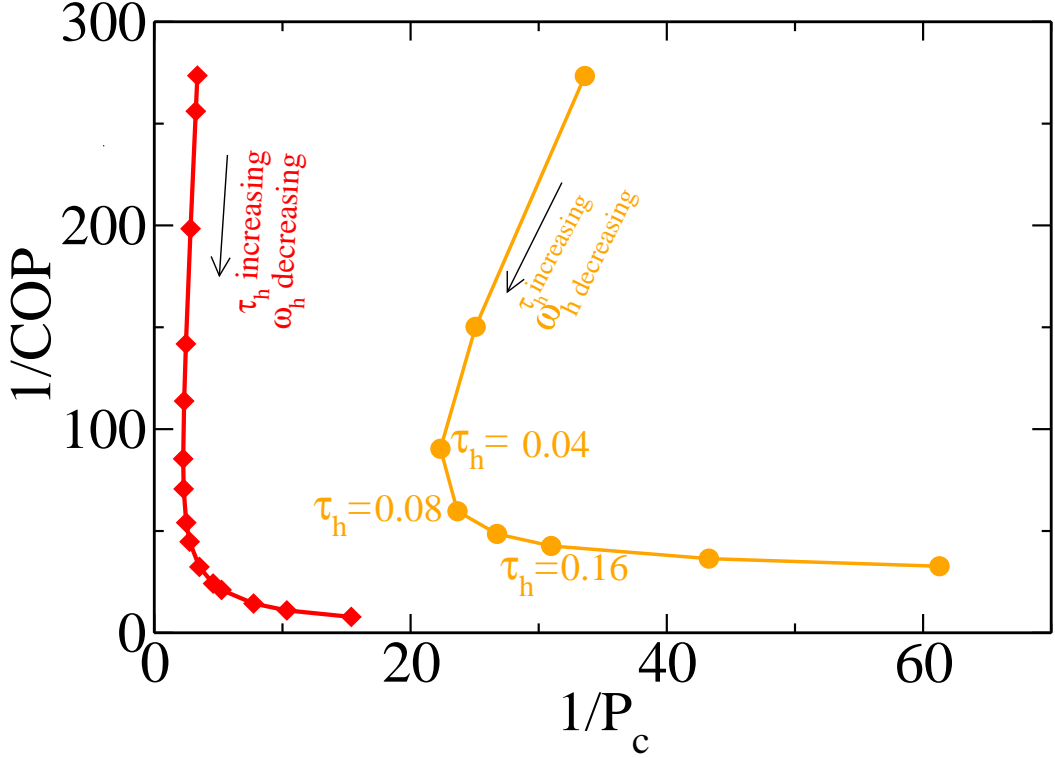}}
\vspace{1.1cm} 
\caption{(Color online)
The $ \bf 1/COP \rm $ as a function of inverse cooling power, $ 1/{\cal P}_c $.
The data correspond to the conditions of approximation
$Q_c^{appr2}$, Eq. (\ref{eq:Qcsecappr2}). The data for both figures are:
$J=2.5,T_h=10,T_c=9,~\omega_c=2.5$, 
$\tau_c=0.008375,~\tau_{ch}=\tau_{hc}=0.065625$
and also the condition $(\omega_h~\tau_h=6.252)$.
 Besides that $\omega_h=625.2$
is the largest  $\omega_h$ value for both plots.
The differences between the plots are the ~$\kappa$ values.
For the orange circle figure 
$\kappa_c^{\downarrow}=0.328,~\kappa_h^{\downarrow}=0.36$,
where three points correspond to the three cycles
of  Fig. \ref{fig:threehot }, whose $\tau_h$ values are
denoted on the figure. The largest $\tau_h$ is $0.32$.
For the red diamond plot 
~$\kappa_h^{\downarrow}=3.6$, and $\kappa_c^{\downarrow}=3.28$.
For this figure the largest $\tau_h$ is $0.5157905$.}
\label{fig:threeappr2}             
\end{figure}

To compute the approximation for ${\cal S}^u$, according to 
Eq. (\ref{apprentr1}), we use ${\cal Q}_C$ from Eq. (\ref{eq:Qcsecappr1b}).
${\cal Q}_h$ requires an additional approximation. To first order  ${\cal Q}_h~=~-{\cal W}^{on}$, 
when ${\cal W}^{on}$ is given by Eq. (\ref{eq:workfirstor}).
therefore ${\cal S}^u$ becomes:
\begin{equation}
{\cal S}^u ~\approx-
\left( \frac{ \alpha  \tau_h^2(\Omega_c ss
 E_h^{eq} \Gamma_h - \Omega_h^2 E_c^{eq}  cc
+E_c^{eq} \alpha \Omega_h^2)}{
(\alpha^2-2 \alpha cc +1)T_c }+\frac{\tau_h E_h^{eq} \Gamma_h}{T_h}
\right)
\approx- \frac{\tau_h E_h^{eq} \Gamma_h}{T_h}
\label{apprentr2}      
\end{equation}   
 Eq. (\ref{apprentr2}) shows that most of the entropy production is generated 
on the hot {\em isomagnetic} segment.
Fig.   \ref{fig:threeappr2}   present the dependence
of the $\bf 1/COP \rm$ on ${1/\cal P}_c$, for the subfamily of cycles
of Eq. (\ref{eq:Qcsecappr2}).  The cycles are chosen with the  condition $\omega_h \tau_h=constant$,
for two different heat transport coefficients to the bath. This analysis in inspired by the 
studies of Gordon et al. \cite{gordon94,gordon97}.
where a universal behavior of optimal cycles with preassigned cycle times was observed.
For the sudden cycles a continuous behavior is only local, nevertheless for the particular family 
by choosing $\omega_h \tau_h=constant$ we found a similar behavior.

\section{The Influence of Cooling  }
\label{sec:coolinginf}

\begin{figure}[tb]
\vspace{-0.66cm}
\hspace{10.cm} 
\center{\includegraphics[height=7cm]{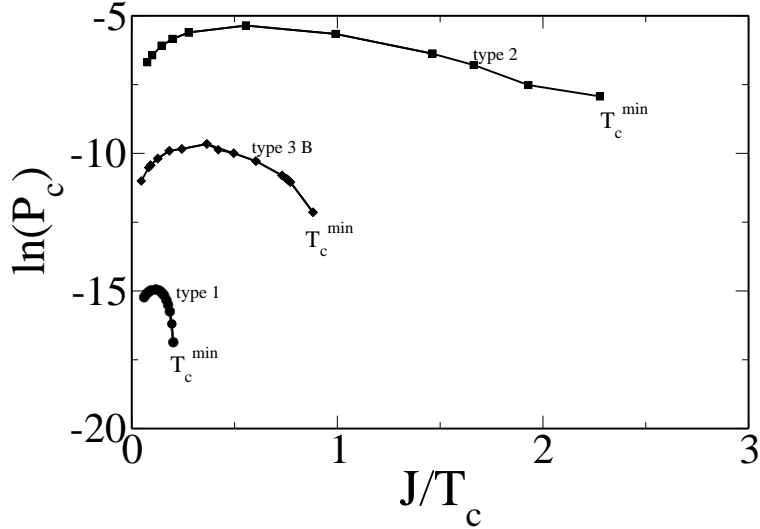}}
\vspace{1.cm}  
\caption{
$\ln({\cal P}_c)$ as a function of $J/T_c$ for the three classes of sudden refrigerators. 
The cycle parameters for type 1  correspond to Fig. 1 where $T_c/T_h$ was kept constant and $T_c$ was varied.
The cycle parameters of type 2 correspond to Fig. 2  where $T_c/T_h$ was kept constant and $T_c$ was varied.
The cycle parameters of type 3b correspond to Fig. 7  where $T_c/T_h$ was kept constant and $T_c$ was varied.  
}
\label{fig:PcvsJpTc}            
\end{figure}  

The cooling power ${\cal P}_c ={\cal Q}_c/\tau $ as a function
of the cold bath temperature $J/T_c$, is shown in  Fig. \ref {fig:PcvsJpTc}. 
A similar plot for the \em regular \rm cycles can be found in ref. \cite{kosloff10}. 
Common to both cases is a minimum temperature beyond which the cycle cannot cool any longer.
This minimum temperature  is obtained when the occupation of the exited level after 
the demagnetization segment is larger than the equilibrium value at the cold bath. 
When $T_c$ is increased the cooling power 
increases exponentially with the equilibrium occupation at $T_c$.  
This minimum temperature is different for the different classes of sudden refrigerators where type 2 
outperforms the other classes in both cooling power and minimum temperature.
In addition a maximum cooling rate  at $T_c > J$ is observed, for all the three classes.
This is in contrast to a monotonic decrease in ${\cal P}_c$ in the {\em regular cycles} \cite{kosloff10,feldmann2010}.
The maximum can be attributed to the inability to dissipate at the hot bath in a very short time the heat at the hot bath. This is a characteristic of the global behavior of the sudden cycles.

The maximum in the cooling rate is also reflected in the approximation Eq. (\ref{eq:Qcsecappr1b}) where 
$\Omega_c \sim J$.
Therefore there exists a positive root of $ x=(J/2T_c)$ in the equation:   
\begin{equation}
\frac{\partial \ln \left( {\cal Q}_c^{appr1b}/\tau \right)}{\partial 
(J/2T_c)} ~\equiv ~\frac{1}{{\cal Q}_c^{appr1b}} \frac{\partial \left( {\cal Q}_c^{appr1b}
\right) }{\partial (J/2T_c)}~=~0
\label{posmaximum}    
\end{equation}   
The derivation is constrained by the fact  that
as  the temperature of the cold bath is varied,  also
proportionally the temperature of the hot bath changes. Writing
$T_c/T_h=C^{T}$, therefore not only $E_c^{eq}$ and $\Gamma_c$ 
are dependent on $T_c$, but
also $E_h^{eq}$ and $\Gamma_h$. 
After quite tedious computation, neglecting terms by order of
magnitude considerations, and taking into account that
in the case  $\Omega_c~\sim J$ one obtains:
\begin{equation}
\frac{1}{{\cal Q}_c^{appr1b}} \frac{\partial \left( {\cal Q}_c^{appr1b}
\right) }{\partial (J/2T_c)}~ =~
2 \kappa_c^{\downarrow} \tau_c \exp^{-2x}+\frac{1}{2J\sinh(2x)}-4 
\kappa_c^{\downarrow} \tau_c \exp^{-2x}~=~0~~,
\label{finalf1}    
\end{equation} 
which leads to:
\begin{equation}
\left( \frac{J}{2 T_c} \right)^{max}=- \frac {1}{4} \ln\left(1-\frac {1}{2J\kappa_c^{\downarrow}\tau_c}\right)
\label{finalf3}     
\end{equation} 
The maximum in  Eq. (\ref{finalf3}) is quite delicate,
this comes about since the constraints on the parameters are very restrictive.
In addition Eq. (\ref{finalf3}) is independent of the ratio $C^{T}$, as well as
other parameters of the hot segment of the cycle. On the other hand,
$(J/2T_c)^{max}$ depends on all the parameters of the cold segment. 
This suggests that the the maximum point of ${\cal P}_c$ could be found also for the case 
where the hot bath has constant temperature.

\section{Discussion}
\label{sec:discussion}
One can say, that practicing science involves making order in
seemingly disorder. The ensemble of the \em sudden \rm cycles is an 
extreme example of that statement.
For example, we saw both in Fig (\ref{fig:seepisland}) and
Fig (\ref{fig:concconv}) that changing the time allocations 
continuously results with a large number of discontinuities.
Therefore continuity, one of the main aids in research, doesn't
help in the case of \em sudden \rm cycles. Also, when one changes
slightly the bath temperatures, or transition probabilities,
or the field values might result with $P_c~<~0$.
One can try to optimize $P_c$ as a function of time allocation,
a standard procedure for \em regular \rm cycles, generally
doesn't work for  \em sudden \rm cycles. The result in most 
cases will be, that while continuously changing the time allocations 
- a jump will occur from refrigerator to non refrigerator. 

The refrigerators studied belong to the family of  four stroke Otto refrigerators \cite{feldmann00,rezek09,feldmann2010,salamon11}.
The working fluid composed from an ensemble of spin pairs, which is a simplified model of 
a working medium composed of magnetic salts.
For this model the dynamics is described by the  equation of motion for the thermodynamical observables.
The present study focuses on refrigerators
with cycle times shorter or much shorter than the internal
time scale of the working fluid. As a result the different segments of the Otto cycle
become interconnected.
This characteristic results in the density operator $\Op \rho$ deviating  
from the typical diagonal  form in the energy representation. 
As a consequence, the energy entropy was always much larger
than the Von Neumann entropy. 
These cycles  termed \em sudden \rm cycles,
settle to a limit cycle, typically after a large number of iterations. 
The state of this limit cycle is the eigenvector of the cycle's
global propagator with eigenvalue one. This property is exploited to study the performance 
of the cycles using a vector space of a closed set of operators which is sufficient
to represent the density operator of the limit cycle.

The total cycle is analyzed through segment propagators which map the vector space of operators.
Our classification scheme of families of sudden cycles is based on analytical approximate expressions 
for the propagators on each segment. 
These segment propagators were then synthesized to
global propagators. Having done that, we computed the approximate eigenvectors 
with an  eigenvalue one for each limit cycle. 
These approximate solution demonstrate the global property of the cycle.
We find a special continuous subset of cycles whose behavior 
is similar to the universal plot of Gordon et al. 
\cite{{gordon94},{gordon97}}. 
It was shown in the study that the entropy production  ${\cal S}^u$
is generated on the boundary of the hot bath.
Finally, the \em sudden \rm cycles possess a maximum cooling power as a function of $J/T_c$
as opposed to the \em regular \rm cycles where we demonstrated a monotonic decrease of ${\cal P}_c$.

Finite time thermodynamics has been devoted to the study of systems far from equilibrium. 
For example the  energy distribution of the working medium was not in the thermal Gibbs state.
The present study is characterized by coherence, a deviation from a Gibbs state because 
the state does not commute with the stationary Hamiltonian.

\section*{Acknowledgments} 

The study was supported by the Israel Science Foundation.
We  thank Amikam Levi, Yair Rezek, Gershon Kuritzky, Robert Alicki and David Gelbwaser
for stimulating discussions.

\appendix
\section{Thermodynamical Relations.}
\label{sec:thermodyn}

In order to fulfill the second law of thermodynamics, the maximal efficiency,
$ \eta^{max} $,  of a heat engine with working fluid of two coupled
spins(with the corresponding Carnot efficiency relation) is
Cf. \cite{kosloff03}:

\begin{equation}
\eta^{max}~=~1~-~ \frac  {\Omega_c} {\Omega_h}~~<~1~-~   \frac {\omega_c}
{\omega_h}~~ ~<~~ 1~~-~~\frac {T_c}{T_h}
\label{eq:heatineq} 
\end{equation}  

For the reverse operation as a refrigerator with the same 
working fluid, the basic inequality must change its direction, and
as a consequence from Eq. \ref{eq:heatineq}, the constraint on
the minimum cold bath temperature $T_c$ will be

\begin{equation}
 {T_c}~~\ge~ ~\frac  {\Omega_c} {\Omega_h} T_h ~\ge~\frac  {J} {\Omega_h} T_h
\label{eq:basicineq} 
\end{equation}

\begin{equation}
 {\bf COP \rm}^{Carnot}~=~\frac {T_c} {T_h - T_c} ~~~;~~~{\bf COP \rm}^{Otto}~=~\frac {\Omega_c} 
{\Omega_h - \Omega_c}
\label{eq:basicCOPs} 
\end{equation}

\newpage

\bibliography{dephc1}

\end{document}